# Near ambient pressure photoelectron spectro-microscopy: from gas-solid interface to *operando* devices

**Matteo Amati[1], Luca Gregoratti[1], Patrick Zeller[1, +], Mark Greiner[2], Mattia Scardamaglia[3], Benjamin Junker[4], Tamara Ruß[4], Udo Weimar[4], Nicolae Barsan[4], Marco Favaro[5], Abdulaziz Alharbi[6] , Ingvild J.T. Jensen[7], Ayaz Ali[7], Branson D. Belle[7]**

[1] Elettra–Sincrotrone Trieste, SS 14 - km 163.5, 34149 Basovizza, Trieste, Italy
[2] Max-Planck Institute for Chemical Energy Conversion Department of Heterogeneous Reactions / Surface Analysis Group, 45470 Mülheim an der Ruhr, Germany
[3] MAX IV Laboratory, Lund University, Box 118, 22100 Lund, Sweden
[4] Institute of Physical and Theoretical Chemistry (IPTC) and Center for Light-Matter Interaction, Sensors & Analytics (LISA+), University of Tuebingen, Auf der Morgenstelle 15, D-72076 Tuebingen, Germany
[5] Helmholtz-Zentrum Berlin für Materialien und Energie, Hahn-Meitner-Platz 1, D-14109 Berlin, Germany
[6] King Abdulaziz City for Science and Technology (KACST), P. O. Box 6086, Riyadh, 11442, Saudi Arabia
[7] SINTEF, Forskningsveien 1, 0373 Oslo, Norway

+ Current position:
Helmholtz-Zentrum Berlin für Materialien und Energie GmbH, BESSY II, Albert-Einstein-Straße 15, 12489 Berlin, Germany
Fritz-Haber-Institut der Max Planck Gesellschaft, Dept. Inorganic Chemistry, Faradayweg 4-6, 14195 Berlin, Germany

E-mail: matteo.amati@elettra.eu

## Abstract

Near Ambient Pressure Scanning Photoelectron Microscopy adds to the widely used photoemission spectroscopy and its chemically selective capability two key features: (i) the possibility to chemically analyse samples in a more realistic environmental, gas pressure condition, and (ii) the capability to investigate a system at the relevant spatial scale. To achieve these goals the approach developed at the ESCA Microscopy beamline at the Elettra Synchrotron facility combines the submicron lateral resolution of a Scanning Photoelectron Microscope with a custom designed Near Ambient Pressure Cell where a gas pressure up to 0.1 mbar is confined inside it around the sample. In this manuscript a review of experiments performed with this unique setup will be presented to illustrate its potentiality in both fundamental and applicative research such as the oxidation reactivity and gas sensitivity of metal oxides and semiconductors. In particular the capability to do *operando* experiment with this setup opens the possibility to perform investigations with active devices to properly address the real nature of the studied systems, because it can yield to more conclusive results when microscopy and spectroscopy are simultaneously combined in a single technique.

## 1. Introduction

The investigation of solid/gas interfaces requires dedicated techniques, and X-ray Photoelectron Spectroscopy (XPS) is one of the most important and solid tools for this kind of characterization because of its surface and chemical sensitivity [1]. Due to technical limitations, XPS instrumentation was originally developed to work in ultra-high vacuum (UHV) conditions using homogeneous samples, its use at more realistic environments requires higher pressures and the study of heterogeneous materials with complex structures, morphologies and interfaces at the smaller scales. These discrepancies are known as the material and pressure gaps.

For instance one of the most studied model catalysts, the Ru(0001) surface, has been investigated for decades with UHV based techniques exploring its attitude to oxidize CO. Only when more realistic pressure environments could be used in the characterization it was found that the low-pressure surface oxide phases investigated in UHV were not the main players in the real catalytic conditions where other phases are formed [2,3]. On the other hand, modern technologies for the production and deposition of catalysts and materials in general do not produce homogeneous samples but composite ones often in the form of micro- or nano-sized particles/clusters with specific structures [4,5] requiring spatially resolved approaches. Therefore revealing the complexity of the processes occurring at the surface/gas interfaces under working conditions, i.e. *in-situ* or *operando*, and at the relevant scale is of paramount importance to understand and optimize the involved process.

To fully exploit the potentials of XPS the above mentioned material and pressure gaps have to be addressed by: (i) adding the possibility of sub-micron spatial resolution, (ii) extending the measurement conditions to pressures higher than UHV. In the past decades solutions to separately address these limitations have been developed, but to design and realize technologies capable of simultaneously overcoming both is extremely challenging.

The synchrotron-based Scanning PhotoElectron Microscopy (SPEM) combines XPS with spatial resolution; it is a powerful technique to image and probe chemical and electronic properties of micro- and nano-structured samples [6,7]. SPEM uses a direct approach to characterize materials at the submicron scale, i.e. to illuminate samples with a sub-micrometric X-ray probe while raster-scanning the sample with respect to it. The SPEM hosted at the ESCAmicroscopy beamline at the Elettra synchrotron research laboratory [8] has been operated since 1996, performing experiments on various types of nanostructured surfaces and interfaces, shedding light on the related phenomena occurring at mesoscopic length scales [9].

The need of a UHV environment in conventional XPS setups is mainly due to the short mean free path of the photoelectrons with kinetic energies below 1500 eV at high gas pressure. These electrons, indeed, suffer from scattering by the molecules in the gas environment. Another reason resides in the presence of high voltages in the electron detector setup which may generate discharges at higher pressures. The state of the art approach to perform XPS at the mbar range, defined as near ambient pressure XPS (NAP-XPS), consists in confining the high pressure only at the sample region, shortening as much as possible the path of photoelectrons in the high pressure by placing the electron analyser nozzle close to the sample area, such distance being typically lower than 1 mm, and differentially pumping the electron analyser lenses stages to create a high vacuum regime not affecting anymore the travel of the photoelectrons. All these conditions are satisfied by the so called differentially pumped Hemispherical Electron Analysers (HEA) [10–14].

The implementation of differentially pumped HEA in SPEM is not straightforward, mainly, because both the analyser and the X-ray optics setup have to be positioned at short distance in front of the sample surface occupying the same space simultaneously. The Near Ambient Pressure Cell (NAP-Cell) developed at the ESCAmicroscopy beamline [15–17] encapsulates the sample within a closed volume and keeps the high pressure locally confined. Such decoupling allows using conventional electron analyser and X-ray optics systems as in UHV setups and therefore offers an alternative solution to the classical near-ambient arrangement.

In the following we will present, together with a detailed desription of the SPEM (section 2) and the NAP-Cell (section 3), some recent results obtained by using the combination of NAP-Cell and SPEM, in order to provide an overview of the capability of this approach in the analysis of gas/solid interfaces and highlight the possibilities in understanding phenomena at the submicron scale at *in-situ* and *operando* conditions.

The studies we propose in this paper will be presented with a progressive increase in the setup complexity to better introduce the different possibility of the combination of SPEM and NAP-Cell in different field. The examples consists in:

(i) fundamental oxidation studies of polycrystalline Ni and Cu foils (section 4 and 5 respectively), where the activity of specific structures will be addressed;

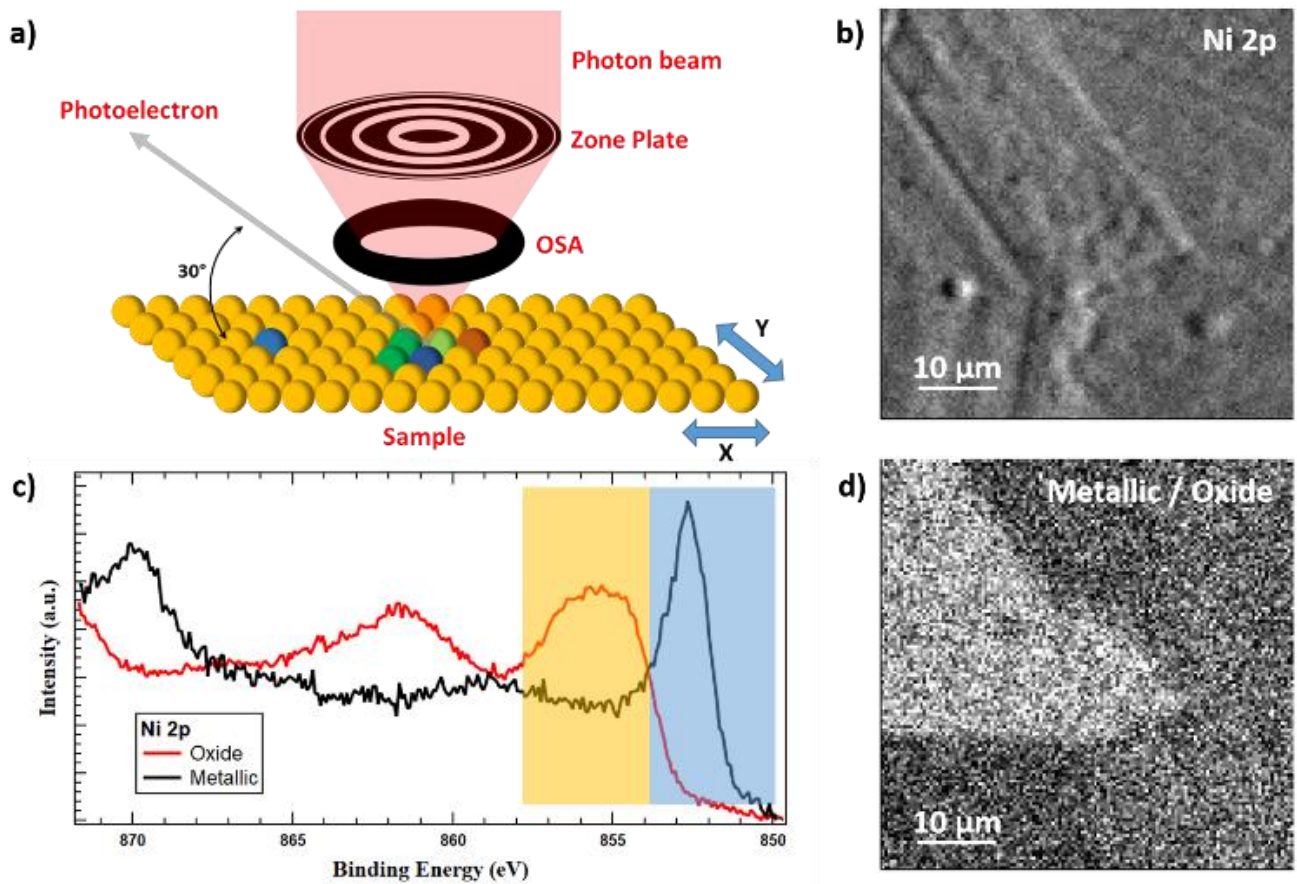

*Figure 1: (a) Sketch of the SPEM setup. (b) Raw photoelectron image acquired at Ni 2p binding energy, 854 eV, with an energy window of +/- 4 eV. (c) Reference Ni 2p photoelectron spectra for metallic and oxidised nickel. The coloured bars in (c) represent the energy window used for the acquisition of map (b). (d) Chemical map extracted from (b) by dividing the two sub maps obtained by selecting the two energy regions showed in (c) by the coloured bars. Brighter points correspond to a higher content of metallic Ni.*

(ii) a more complex system consisting of graphene on copper (section 6) where the oxidation propagation underneath graphene flakes will be studied;

(iii) gas sensors active devices based on semiconducting metal oxides (section 7) and a single 2D $MoS_2$ flake (section 8) where inside the NAP-Cell not only sample temperature and gas pressure are controlled, but also potentials have been applied to the systems and electrical responses have been recorded during measurements, i.e close to *operando* in NAP-Cell more realistic conditions.

Each example will be presented with a brief introduction to the corresponding research field and topic.

## 2. Scanning photoemission microscopy basic concepts

To produce a sub-micrometric probe the incoming X-ray beam is focused by means of lithographically made Fresnel Zone Plates (ZP) [18] which generate a large number of diffracted orders each having a specific intensity and focal position. In the SPEM only one order must be used to create the focused X-ray spot. To do so, the first diffraction order is selected and the higher ones are blocked by an additional pinhole, the so called Order Selecting Aperture (OSA). The straight radiation passing through the rings, defined as the zeroth order, is stopped by a metallic X-ray absorbing layer, called central stopper, lithographically grown on the ZP centre.

The typical set-up used at ESCAmicroscopy consists of a 200-250 µm diameter ZP lens with 50-100 nm outermost ring and a 80 µm central stopper, and a 75 µm wide OSA. With the photon energies available at the beamline, in the 400-1200 eV range, this focusing system is capable to de-magnify the photon beam down to a spot of 130-180 nm diameter providing a focal length in the 5-15 mm range, with both the spot dimension and focal length being function of the photon energy [18].

As sketched in Figure 1, samples are placed in the focal point of the optics that work in normal X-ray incidence. Sample is then raster scanned, in the plane orthogonal to the beam, by two sets of motors. The first one consists of stepper motors designed for large movements, up to several mm, with a minimum step movement of 1 µm, and the second one is a piezoelectric stage capable of 100x100 µm scanning range

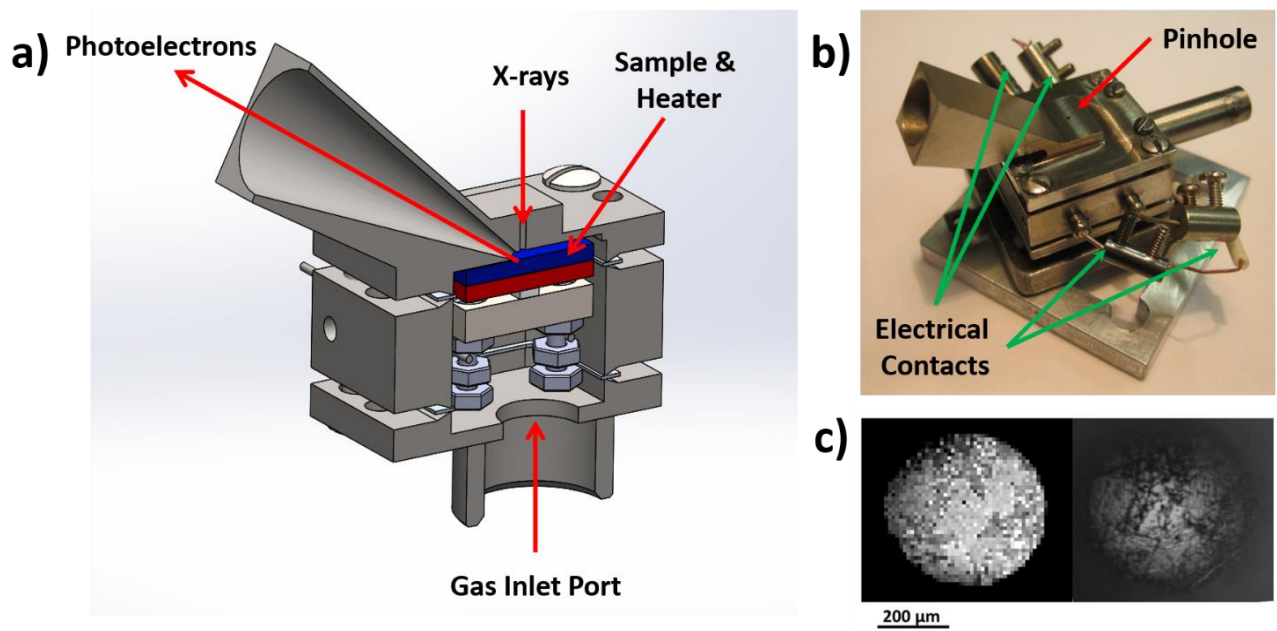

*Figure 2: (a) Sketch of the NAP-Cell setup. (b) Photography of the NAP-Cell with all four electrical feedthrough and the X-ray pinhole highlighted. (c) Comparison between a SPEM image, on the left, and optical image, on the right, of the field of view available for SPEM measurements on the sample.*

with a minimum step movement of 5 nm for high spatial resolution imaging of samples.

Due to the geometric restrictions imposed by the short focal length, the photoelectrons detectable by the HEA have a take-off-angle with respect to the sample surface of 30°, Figure 1, and the HEA maximum acceptance angle is ± 10°. This configuration strongly enhances the surface sensitivity of the instrument. The HEA equips a delay line electron detector that is binned to 48-channels [19].

The SPEM can operate in two modes: (i) imaging spectro-microscopy and (ii) micro-spot spectroscopy. The imaging mode maps the lateral distribution of elements or chemical states by collecting photoelectrons within a selected kinetic energy window while scanning the specimen with respect to the microprobe. The micro-spot mode is identical to a conventional XPS spectra, but it is measured from selected submicron spot on the sample.

In Figure 1(b) an example of a photoemission image, acquired at a partially oxidized Nickel foil using the Ni 2p signal, is reported. The grey scale is proportional to the photoelectron intensity recorded by the HEA in that particular point. The contrast in the image is mainly generated by surface topography rather than by chemical variations, with the intensity in each point reflecting its orientation with respect to the HEA; in particular a region oriented toward the HEA appears brighter due to more normal emission. Nevertheless, the 48 channels electron detector allows to highlight in the image the lateral distribution of the different chemical states and to remove the topographic contributions. At each scanned position the corresponding 48 points XPS spectrum defined by the selected window energy is stored generating an array of 48 maps each showing the distribution of a specific photoelectron energy. By properly choosing the map acquisition energy and analysing each of these spectra it is possible to create maps corresponding to specific spectra features. For example in Figure 1 (c) two energy regions corresponding to metal, 850-854 eV, and oxide nickel, 854-858 eV are highlighted. The ratio, point by point, of the integrals of the 48 points spectra from these two energy regions generates the map showed in Figure 1 (d). Here, the greyscale represents the metal to oxide ratio and thus identifies local chemical differences. In addition, the topography contribution is cancelled [19].

## 3. Near Ambient Pressure Cell

Higher pressures at the sample can be achieved by a special designed cell, that contains the sample and decouples sample environment from the rest of the main chamber [15–17]. The top plate of the NAP-Cell contains two 450-500 μm diameter pinholes whose impedance creates the needed pressure drop between the inside volume of the cell and the main chamber. One pinhole is perpendicular to the sample and is used by the incoming focused X-ray beam. The second pinhole for the emitted photoelectrons is oriented at 30° and has a conical shape to properly fit the HEA acceptance angle. The sample inside the NAP-Cell (Figure 2) is located 650 μm far away from the surface of the inner body of the cell. This position allows the sample to be at the intersection point of the axis of the two pinholes. A flexible pumping/dosing line is connected

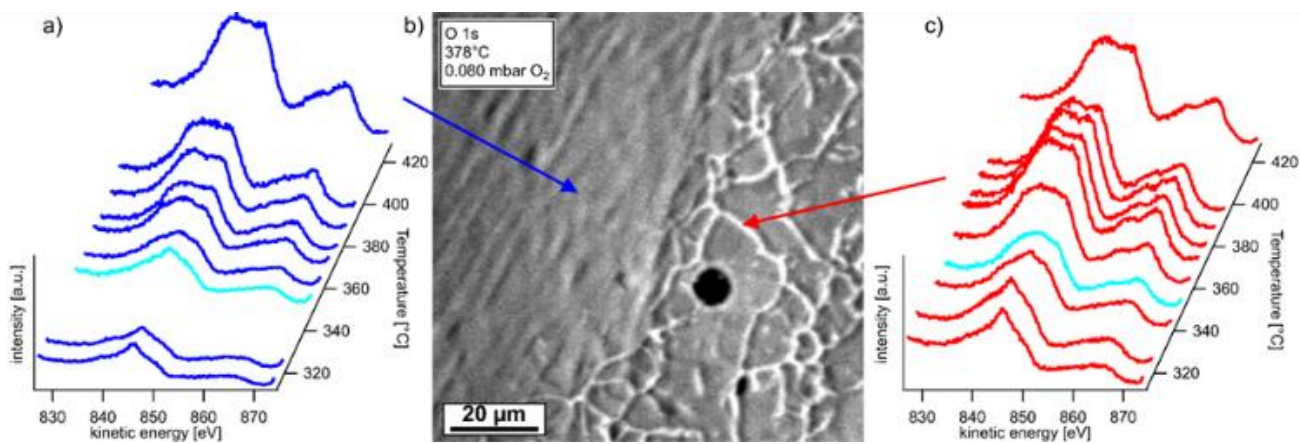

*Figure 3: Oxidation of a polycrystalline Ni foil. The Ni LMM spectra were taken (a) at flat Ni regions and (c) localized at a grain boundary. The spectra acquired at 350°C (marked in light blue) show the largest deviation in peak shape. (b) The map taken at the O 1s core level shows the investigated area (BE centred at 530.5 eV; PE = 40 eV).*

to the cell to control the environment without affecting any sample movement.

The compact design of the NAP-Cell allows to mount it on a standard SPEM sample holder and to fit in the limited space between the sample and the focusing optics, making feasible to operate the SPEM in the same operation modes described above. The pressure inside the cell can be raised up to 0.1 mbar while the pressure in the main chamber remains below $1 \times 10^{-5}$ mbar, which is the safety limit for the SPEM system without the necessity of any differential pumping system.

A ceramic heater is located behind the sample for controlling its temperature in the range 300–1073 K. Samples are electrically insulated from the body of the cell, having four independent electrical contacts for potential and current applications, see Figure 2 (b).

With this design a 450 µm wide sample field of view can be achieved, as shown in Figure 2 (c). Moreover, the gas composition of a chemical reaction can be determined by a mass spectrometer located in the main chamber.

## 4. Oxidation of polycrystalline Nickel

Catalysts used in heterogenous catalysis are typically complex structures of active materials dispersed on a support and with addition of promotors [20,21]. This complexity entails often a chemical inhomogeneity that could only be tackled by spatially resolved techniques [21]. In addition, these investigations have to be done at *in situ/operando* conditions because the phase of interest may not exist in initial or post mortem studies [21]. Up to now there are only a few techniques that fulfill these requirements [15,22–24].

The investigations were performed using a 0.25 mm thick polycrystalline nickel foil (Advent Research Materials Ltd, purity 99.99%). After mechanical polishing the sample was cleaned by repeated cycles of sputtering (1.5 keV, 30-60 min) and annealing (~820°C, 60-90 min). After transferring the Ni foil into the NAP cell and right before the experiment, additional cleaning cycles of oxidation (360°C, 0.07 mbar $O_2$) and reduction (420°C, 0.18 mbar $H_2$) were performed. The photon energy was set to 974.5 eV.

### 4.1 Results and discussion

At first, the morphology of the polycrystalline Ni-foil was investigated using the imaging capabilities of SPEM. A representative image acquired from this sample using the O 1s signal is shown in Figure 3 (b). As described in sections 2 and 3, the XPS maps acquired by SPEM are a superimposition of chemical and topographic information, while in the present case topography is dominating. According to topography the shown region can be divided into several parts. At the left side of the image the surface is mostly flat with only small ripples that are typical of rolled metal foils. On the other hand, at the right side of the map there is a strong contrast with pronounced, bright lines. To the right of each bright line is an additional faint black line. Due to the grazing analyzer position these adjacent lines can be identified as trenches. Most likely, the origin of these trenches are bulk grain boundaries that result in a strong curvature of the material when reaching the surface. Following this, the right image part can be interpreted as the result of the presence of many small grains with the size of few micrometers separated by grain boundaries. Therefore, the bright features will be called grain boundaries in the following.

These variations in topography may strongly affect the local chemistry of the nickel foil. As a probe for this the oxidation of the Ni was chosen and its progress was monitored at selected regions, i.e. at single grains, at flat regions or at the strong curvature of the grain boundaries. To do so, 0.08 mbar of oxygen was introduced and sample temperature was stepwise increased from 315 to 430 °C. At each temperature step, spectra at selected points of interest were acquired. The map shown in Figure 1 b) was taken from this oxidation series.

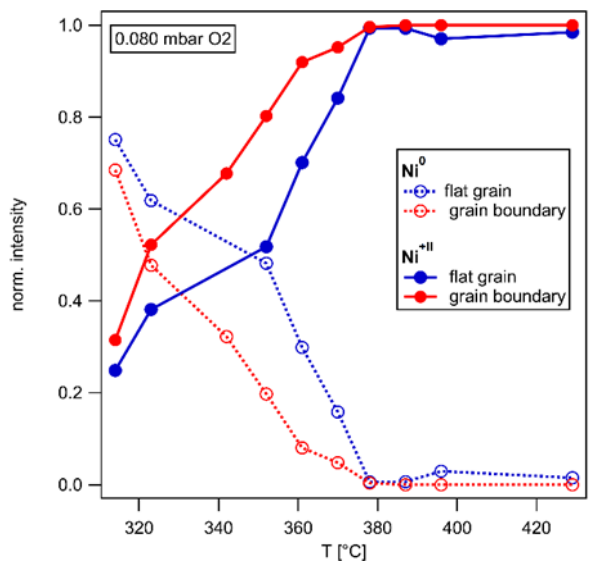

*Figure 4: Ni oxidation state during the oxidation. The oxidation state was obtained by a linear combination fit of the data shown in Figure 3 (a) and (c).*

Ni LMM Auger spectra series acquired at the flat region and at one of the grain boundaries are provided in Figure 3 (a) and (c), respectively. In both series, there is between 315 and 380 °C a strong change in the Auger peak shape. As it is highly sensitive to the Ni oxidation state [25], this can be attributed to a transition from mostly metallic towards oxidized Ni. A closer inspection of the data reveals that the change in peak shape happens at lower temperatures at the grain boundary compared to the flat Ni, best visible at the spectra acquired at 350 °C (highlighted in light blue). For better visibility the oxidation state of the Ni was extracted from the Auger spectra by a linear combination fitting using reference peak shapes. The results are provided in Figure 4 and confirm the differences in Ni oxidation states at the different regions. In both cases there is between 315 and 370 °C an almost linear increase of the fraction of oxidized nickel and eventually a complete oxidation around 380 °C. The corresponding curve for the grain boundaries is shifted by about 20 °C towards lower temperatures compared to the flat regions, i.e. the oxidation of grains boundaries occurs at a 20 °C lower temperature. This implies a higher reactivity of the sample towards oxidation at the grain boundaries which is in line with their expected behavior [23,26,27] due to their high number of defects, high atomic step density and the fact that grain boundaries can act as a diffusion path of adsorbed oxygen into the bulk [28]. Although the density and spatial dimension of the grain boundaries are quite high in the shown region, the bright lines contribute only about 7 % to the area of the shown map. Therefore, any effect of the grain boundaries would be hardly visible using averaging techniques rather than spatially resolved.

This preliminary study shows the feasibility to direct address the surface chemistry at selected topographical features using the combination of SPEM and NAP conditions. Next step will be to proceed with a heterogeneously catalyzed reaction and investigate these topographical features of the catalyst at *operando* conditions which is very difficult with other techniques. These features could be active sites of the catalysts and therefore investigating their local chemistry would help understanding their working principle [20,23,26,27].

## 5. Formation of metastable phases during oxidation of polycrystalline Copper

Metal oxidation is an interesting topic, considering its importance in high temperature corrosion, as well as general solid-state surface reactivity. The knowledge of bulk phase diagrams for oxidation products is highly developed. However, much less well understood is the formation of meta-stable, sub-stoichiometric phases that might only exist on the nano-scale, and for only short durations, but are nonetheless important for the kinetics of the oxidation. There is certainly substantial evidence of the formation of defective, sub-stoichiometric, meta-stable oxides formed during oxidation, but the lack of knowledge in this field is because experimental observation of such phenomena is technically very challenging. In-situ photoemission spectro-microscopy methods have recently made great progress in this direction.

Investigations on sub-stoichiomeric and meta-stable phases benefit greatly from the ability to detect signals representative of chemical state, but from spatial domains on the nanometer scale. The reason why this is important, is that sub-stoichiometry as well as metal coordination geometry influences photoelectron signals, such as XPS. Thus, evidence of sub-stoichiometry can generally be found in electron spectra. However, when meta-stable states are minority species, present simultaneously with the majority bulk thermodynamically stable species, the signal from the meta-stable species can hardly be discerned. For this reason, it is a great asset to such investigations to be able to spatially isolate the signal from the meta-stable species.

A polycrystalline copper foils of 0.1 mm thickness and 99.998% purity from Advent Research Materials Ltd was used for this experiment. Prior to measurements, copper samples were cleaned using cycles of $Ar^+$ ion bombardment polishing followed by annealing in 0.2 mbar high purity (99.999%) $H_2$ at 700–800 °C. The result was a fully reduced surface.

A combination of in-situ SEM and in-situ SPEM was used to examine the oxidation of the Cu foils and in particular the oxide phase transition that is not accessible using traditional UHV spectroscopic methods. The oxidation was performed by exposing the Cu foils to *0.1 mbar $O_2$ at 300°C*. The transition of $Cu_2O$ to CuO, that typically occurs during atmospheric oxidation of Cu, is not accessible at very low $O_2$ pressures. However, using spectro-microscopic methods in the mbar pressure range, one can observe this transition live, and with chemical detail. The SEM image in Figure 5 (a) shows that some sort of intermediate morphology forms during the onset of the phase transformation from $Cu_2O$ to CuO. This intermediate morphology is represented by the light grey feature spreading across the dark-grey surface. The dark grey is $Cu_2O$. The cauliflower-like morphology is CuO (as determined by post-reaction analysis not shown here). However, the middle grey morphology is unknown from SEM images, and does not remain after cooling the sample down and moving to air, preventing post-reaction analysis.

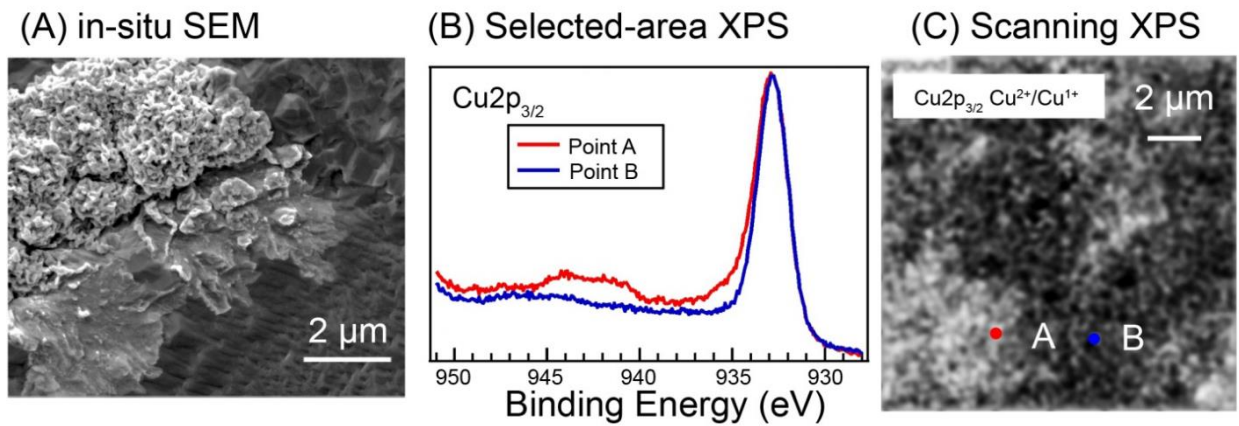

(A) in-situ SEM     (B) Selected-area XPS     (C) Scanning XPS

*Figure 5: (a) In-situ SEM image of $Cu_2O$ while heated in 0.3 mbar $O_2$ at 300°C, showing the initial formation of CuO. The dark grey area is $Cu_2O$, the middle grey area is an intermediate morphology that precedes bulk CuO formation, the bright cauliflower-like morphology is bulk CuO. (b) In-situ microspot Cu $2p_{3/2}$ XPS spectra measured from the growth front of an the CuO phase transition (red) and from the $Cu_2O$ phase (blue), as indicated in the Cu $2p_{3/2}$ $Cu^{+2}$ / $Cu^+$ SPEM map in panel (c) The intensities of the two spectra have been normalized with respect to the peak at 934 eV for clarity. The conditions for the SPEM measurements were 0.1 mbar $O_2$ at 300°C.*

This interesting morphology is somehow related to the initial formation of CuO, and could potentially represent a meta-stable, sub-stoichiometric oxide phase. However, to determine its identity, one requires in-situ spatially resolved spectroscopic methods that are able to characterize chemical state. To this end, in-situ SPEM was applied. The photoemission map in Figure 5 (c) shows the $Cu^{2+}/Cu^{1+}$ contrast; despite the lower spatial resolution regions with similar morphologies as in Figure 5 (a) can be found. Panel (b) in Figure 5 shows Cu $2p_{3/2}$ point spectra from the points A and B indicated in the figure. Spectrum A was measured directly at the edge of a growing CuO front, and shows the growth front to contain a superposition of $Cu^2$ and $Cu^{1+}$ spectra. The $Cu^{2+}$ component is identified by the satellite peaks between 940 and 945 eV, as well as a main peak at 934 eV, whereas the $Cu^{1+}$ component consists of only a main peak at 932.5 eV. Point B was measured in the middle of a $Cu_2O$ phase and shows only evidence of $Cu_2O$.

The ability to have chemical identification on the nanoscopic scale, and under in-situ conditions, is very powerful in that it can yield more conclusive results when microscopy and spectroscopy are combined.

## 6. Graphene barrier towards Cu oxidation

The study and development of anticorrosive coatings is of high technological interest and strategic industrial importance due to the high costs that metal corrosion implicates [29]. Oxidation is an example of high-temperature corrosion that happens when a metal is exposed to an oxidising atmosphere at elevated temperature. This is a typical environment in gas turbines, engines or furnaces, as an example. Oxygen and nitrogen dioxide are typical aggressive gases that can enter in contact with metal parts. $O_2$ is naturally present in air, while $NO_2$ is a common air pollutant, contributing to the formation of smog and acid rain and a product of combustion in car engines. Graphene, an intrinsic 2D material, is impermeable to any molecule and inert to most chemicals [30]. It acts as a barrier for the supporting substrate, being a promising candidate as an anticorrosive coating for metals [31–34]. The case of copper, in particular, is being recently studied in deep since it is a common substrate for graphene growth, other than for its wide use in everyday life. It was shown that the protection offered by graphene is very effective in different reactive environments, as high relative humidity [35,36], hydrogen peroxide [37] high temperature in oxidizing gases [31,32,38]: when the exposed metal areas were already corroded, the region protected by graphene were still unaffected.

SPEM has revealed to be an useful tool to study supported and suspended graphene [39–44], opening the possibility to follow the different kinetics involved in this process at a single flake level.

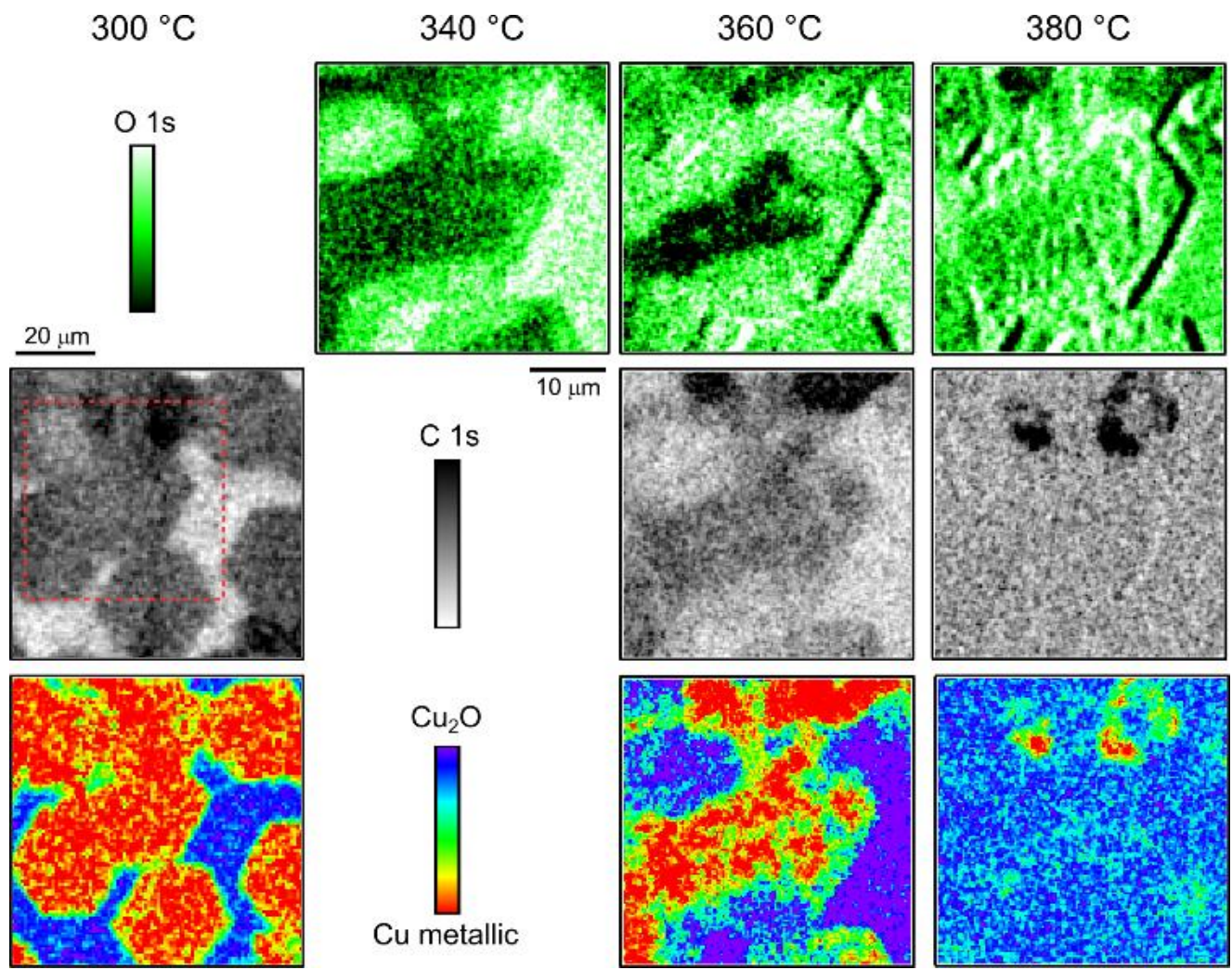

*Figure 6: O 1s (top row), C 1s (middle row) and Cu LMM (bottom row) photoemission maps of graphene/Cu foil at different temperatures. The red dotted square in the 72x72 μm² C 1s map represents the zoom-in of the following 38x38 μm² maps. The Cu LMM Auger maps are have been processed in order to distinguish the contribution of metallic copper and Cu₂O indicated with different colour in the images.*

In this experiment, we studied the oxidation of copper protected by graphene flakes in different gas environments ($O_2$ and $NO_2$ at 0.1 mbar) and at a variable temperature. The graphene flakes are not covering the whole copper surface, thus having available under the same field of view both unprotected Cu areas and graphene-covered Cu. The flakes are mostly single layer of about 20 μm in side, with few multilayer areas. In Figure 6, the 72x72 μm² C 1s map at 300 °C clearly shows graphene flakes with well-defined hexagonal shape, the darker spot on the top of the map indicated the presence of a multilayer region, while the brightest part is the bare Cu substrate. The corresponding map of copper is acquired on the Cu LMM Auger window (centred at a kinetic energy of 918 eV) and the chemical analysis by making the ratio of fingerprint regions is shown here. The map clearly shows that the area uncovered by graphene are already oxidised to $Cu_2O$, the copper native oxide, while the Cu beneath graphene flakes is still metallic [31,45].

When the temperature is increased to 340 °C, the oxygen molecules start to intercalate under the graphene mostly from flakes' edges, as showed in the zoomed 38x38 μm² map in Figure 6. Contextual to the intercalation is the starting of oxidation of the metallic copper underneath. The etching of graphene, however, does not happen at the same time of copper oxidation which indeed is an undercover reaction [31,32].

C 1s spectra (not shown here) measured in different areas are centered at two different binding energies (BE): 284.1 eV as nominal position for Gr [41] and shifted to lower binding energy of about 0.4 eV, in agreement with oxygen-intercalated graphene where the graphene is lifted from the substrate and the interaction is therefore reduced [46]. The spatial distribution of these C moieties is shown in Figure 7 (a) corresponding to the 360 °C condition. Zone A (dark blue) corresponds to areas where the C 1s peak is centred at its original position, zone B (light blue/green) to areas with the shifted peak and zone C (yellow/red) indicates the absence of carbon signal.

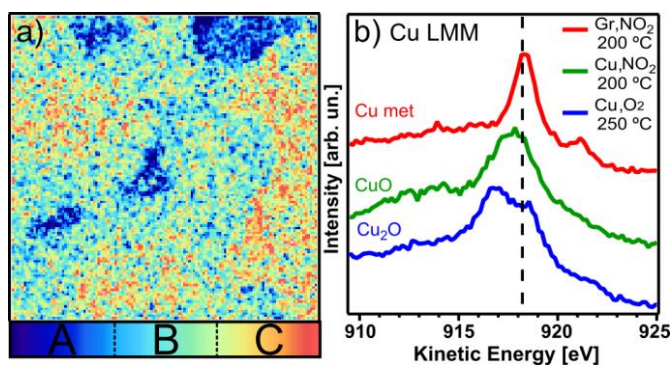

*Figure 7: (a) 38x38 µm² C 1s photoemission map at 360 °C obtained by making the ratio of shifted and pristine C 1s fingerprint regions in the energy window. The different region correspond to pristine (A) and O₂-intercalated (B) graphene and carbon free areas (C). (b) Cu LMM Auger spectra at three different conditions.*

When the temperature is further increased beyond 380 °C the single-layer graphene is etched away. Only small multilayer regions are still present and under them the copper maintains its metallic character: they are visible in the upper part of the maps at 380 °C in Figure 6. The etching of graphene happens through combustion with the oxidising gas and it is facilitated by the change of the copper morphology which, at this stage of oxidation, creates protuberances on the surfaces [31,47]. This is evident in the O 1s map in Figure 6 where the topographical contribution was intentionally not corrected and shows the presence of regions with different height on the sample, represented by sharp dark/bright edges.

When $NO_2$ instead of $O_2$ is used, a higher and faster aggressiveness of the oxidation process is noticed [48]. The formation of the second oxide of copper, CuO, is one of these effects. The lineshape of the Cu LMM Auger spectrum is the best fingerprint to disentangle the different Cu phases [32,45], thanks to the different kinetic energy of the main peak, the chemical contrast maps in Figure 6 were obtained. In Figure 7 (b) the Cu LMM Auger spectra recorded at different temperature and gas composition on the graphene/Cu sample are reported. At 200 °C in $NO_2$ atmosphere, the copper protected by graphene is still in its metallic phase (red spectrum) with the main peak at 918.3 eV, while the bare Cu surface is oxidised to CuO (green spectrum, main peak at 917.9 eV). On the contrary, while exposing to $O_2$, at 250 °C the copper still shows sign of metallic phase together with the formation of $Cu_2O$ at 916.6 eV (blue spectrum). The formation of CuO in oxygen atmosphere is subsequent to the formation of $Cu_2O$ and evolves with the temperature [32].

In summary, with NAP-SPEM we were able to follow the oxygen intercalation and undercover copper oxidation beneath graphene and contextually to establish the mechanism of graphene protection towards copper oxidation in different aggressive environments, namely $O_2$ and $NO_2$ atmosphere, at high temperature.

# 7. Characterization of LaFeO₃ based gas sensors

Semiconducting metal oxides are the basis for the majority of chemoresistive gas sensors that are used in many applications such as detection of flammable and hazardous gases, indoor air quality management or industrial process control [49–52]. While widely used oxides such as $SnO_2$, $WO_3$ or ZnO respond to many target gases, gas sensors based on $LaFeO_3$ (LFO) with perovskite structure exhibit a high sensitivity towards acetylene and ethylene and, more interesting, reveal a temperature dependent intrinsic selectivity to these target gases [53,54]. In order to investigate the detailed mechanisms of the gas–solid interaction, research is mainly focused on *operando* techniques where the sensor's surface can be studied in conditions similar to the intended future application. In this study on LFO based gas sensors in different atmospheres, local variations at the sensor surface under realistic operation conditions are identified via spatially-resolved elemental maps obtained using SPEM. Even if the maximum pressure of 0.1 mbar available in this setup is significantly lower than atmospheric pressure, its effect on the chemistry of the sensor's surface is still detectable.

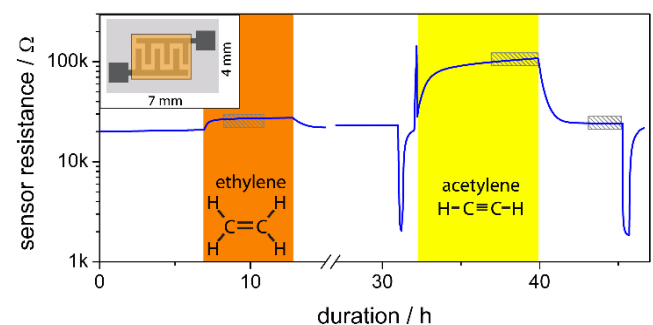

*Figure 8: The sensor resistance during the NAP-Cell experiment in different atmospheres. The exposure to ethylene and acetylene is marked by a coloured background. The grey areas indicate the time when the spectra presented in this study were recorded. The inset shows the layout of the sensor substrate including Pt electrodes and the sensing layer printed on top of it.*

## 7.1 Experimental

The LFO based material was synthesized by a sol-gel process starting from $La(NO_3)_3 \cdot 6H_2O$ and $Fe(NO_3)_3 \cdot 9H_2O$ as described in a previous publication [53]. The obtained gel was calcined at 600 °C for 2 h. The sensing device consists of an alumina platelet (4×7 mm²) with interdigitated Pt electrodes at the surface (electrode gap: 300 µm). The sensitive layer is deposited onto the substrate via screen printing with a thickness of approximately 50 µm (see inset of Figure 8). The sensor was heated to 300 °C using the resistive heater of the NAP-Cell. The atmosphere in the cell consists of dry synthetic air (20.5% $O_2$ in $N_2$) as well as synthetic air with 1 vol.-% of ethylene ($C_2H_4$) or acetylene ($C_2H_2$), respectively, at a total pressure of 0.1 mbar. Synthetic air as well as the gas mixtures were provided by Westfalen AG, and were directly introduced into the NAP cell. The resistance of the sample was monitored with a Keithley 617 electrometer in Ohms mode with one of the electrodes connected to the spectrometer ground. Spectra of various peaks were recorded with a photon energy of 975 eV and a pass energy of 20 eV.

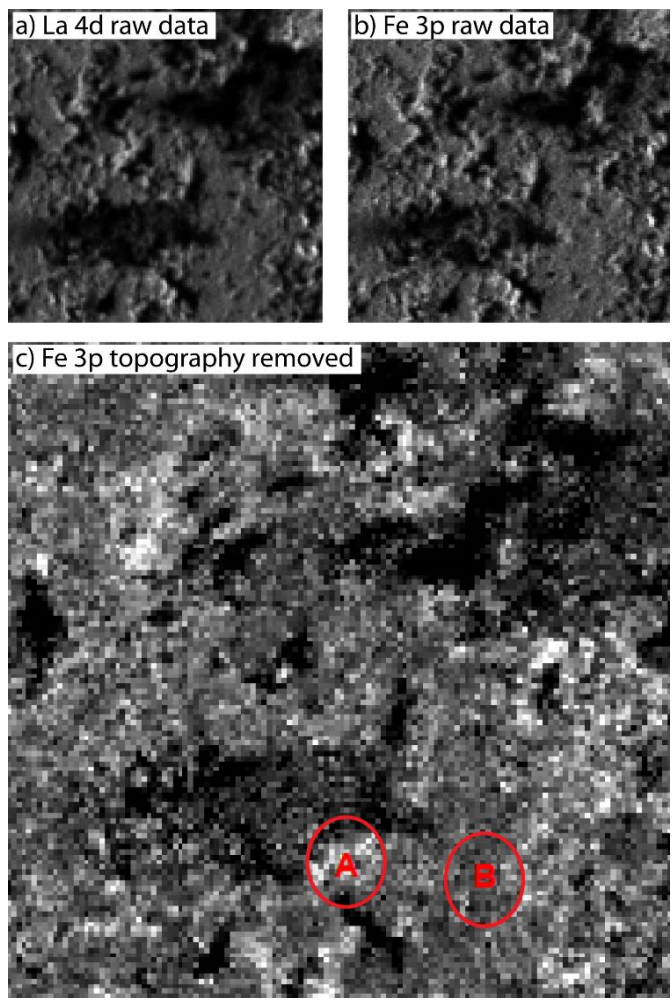

a) La 4d raw data

b) Fe 3p raw data

c) Fe 3p topography removed

*Figure 9: maps of the sample surface (100×100 μm²) recorded at the binding energies of La 4d (upper left panel) and Fe 3p (upper right panel) as well as a map of Fe 3p with the topography removed. The spots marked A and B have been selected for microspot XPS analysis. All maps were recorded during exposure to acetylene*

### 7.2 Results and discussion

Even though the response of the sensor to acetylene and ethylene is significantly higher at 250 or 200 °C, respectively, the study was carried out at 300 °C to avoid the deposition of carbon at the surface during the experiment. In order to compensate for the lower signal, a rather high concentration of hydrocarbons of 1 vol.-% was chosen. The results of the DC resistance measurements are given in Figure 8. As expected for a p-type semiconductor, the extent of the hole accumulation layer decreases and, hence, the resistance of LFO increases during the exposure to reducing gases such as ethylene and acetylene [55,56]. The obvious decrease in resistance around 31 and 45 h, respectively, is due to heating the sample above the temperature of operation in order to remove carbon that was adsorbed at the sample surface while the sharp increase in resistance around 32 hours is due to a decrease of pressure. The sensor response to acetylene is significantly higher than for ethylene which is in line with previous *operando* investigations [53].

The characterization of the surface chemical composition was performed using SPEM, acquiring the Fe 2p and 3p, La 3d and 4d, O 1s, and C 1s core level peaks at a constant photon energy of 975 eV. The maps for La 4d and Fe 3p are given in the upper panels of Figure 9. The contrast is dominated by the topography of the sample. As the contrast is identical in both maps, the material showed excellent homogeneity in terms of elemental composition and no obvious areas with an excess of La or Fe or even separate phases could be identified. The topography can be removed by making ratios of properly chosen regions of the energy window, and an example for Fe 3p is given in the bottom panel of Figure 9, where the spectra region was divided in half and the ratio between this two half is shown. The areas around spots labelled A and B have been identified to be promising candidates for locally recording XPS spectra.

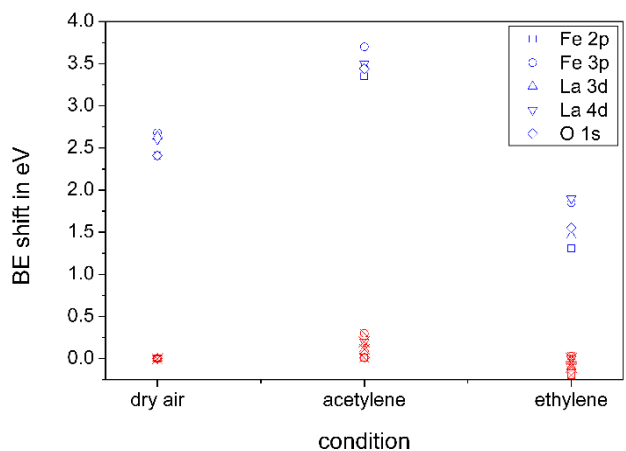

*Figure 10: shift in binding energy of different signals for point A (red, crossed) and point B (blue, open) in different atmospheres relative to point B in dry air.*

A comparison of the metals' spectra, especially their oxidation states and multiplet splitting, didn't reveal any differences neither between point A and B nor in different environmental conditions. The introduction of gases however has a remarkably different effect on the BE of point A and B. For point B the shift in binding energies between different conditions is less than 0.3 eV. In contrast, the binding energies for point A are shifted towards higher binding energies by several eV. In Figure 10 all shifts are plotted relative to the binding energy of point B in dry air. Spectra recorded without the focusing optics, thus averaging over a 75 µm diameter area, not shown here, show the same binding energies as point B and hence the binding energies of point B in dry air are taken as a reference for the shifts shown in Figure 10. One of the possible explanations for the distinct peak shift on spot A may be local charging due to a small volume at the surface of the LFO layer being only loosely connected to the bulk and the grain boundaries acting as resistors whose values are dependent on the gas atmosphere. The electrons released by the reaction of acetylene or ethylene with LFO recombine with holes in the valence band. The decreased conductivity at the grain boundaries could make the area around spot A electrically more isolated and thereby charging intensifies. From evidence data it is evident that the effect of acetylene on the conductivity is higher than the one of ethylene and the same can be observed for the binding energy shift.

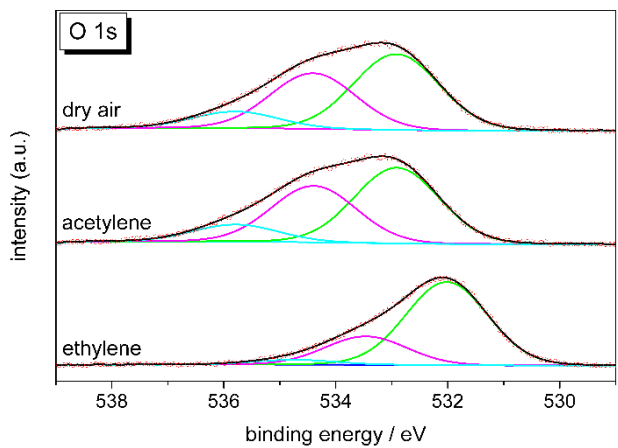

*Figure 11: microspot XPS of the O 1s region on spot A in different atmospheres. The experimental data (circles) have been fitted with three components: Bulk oxygen (green), OH/formate species (magenta) and other adsorbates (cyan). The sum curve of the fit is drawn in black.*

Interestingly, the peak shift in dry air, recorded after the exposure to acetylene, is higher than in the presence of ethylene. These findings as well as the spectra of the O 1s region indicate a permanent change in the surface of the sensing layer. A deconvolution of the of O 1s data was possible using three components with the main one representing the bulk oxygen (green line at lowest binding energy in Figure 11). The second component ($\Delta BE$ = 1.4-1.5 eV) can either be assigned to hydroxyl groups or formate species at the surface and the third peak represents various other surface adsorbates such as carbonates ($\Delta BE \geq 2.4$eV) [57–59]. Unfortunately the interpretation of C 1s spectra could not provide additional information on the origin of these components, since the charging and the generally decreased amount of carbon species on point A do not allow a direct comparison with point B. Moreover, the microspot XPS of O 1s on location A reveal an increased contribution of oxygen containing surface species compared to the bulk component in the spectra recorded subsequent to the ones in ethylene. To a lesser extent the same can be observed on spot B. Since La and Fe didn't reveal any changes in their oxidation states and the O 1s and C 1s spectra between points A and B differ, the conclusion can be drawn that organic surface components formed by the exposure to the target gases play a key role in the reception of the gas sensing process. The identification of these species requires additional spectroscopic investigations and may finally be required to explain the inhomogeneous behaviour of the surface and the origin of the significant shifts in binding energy.

In this study, NAP-SPEM was successfully applied to a gas sensitive material. Although the sample surface was chemically homogeneous, sites with different reactivity towards the target gases ethylene and acetylene were identified. These findings, crucial for an applied device, would not have been revealed with averaging spectroscopic techniques. Further studies of non-homogeneous samples, e.g. heterostructures or materials with additive clusters, will be performed in the future using the same approach delineated here.

# 8. MoS2 monolayers as gas sensors: an operando study

Two dimensional (2D) layered nanomaterials are promising candidates for ultra-sensitive gas sensors due to their high surface to volume ratio, high carrier mobilities and tuneable bandgaps[60,61]. For field effect transistors (FETs) based on 2D layered molybdenum disulphide ($MoS_2$), sensitivities down to parts-per-billion towards $NO_2$ have been demonstrated [60,62–64]. The sensing mechanism is believed to be based on a charge transfer process [65,66] where oxidising (reducing) gases accept (donate) electrons from the channel upon adsorption, leading to an increase (reduction) in the resistance.

## 8.1 Methods

*Device fabrication process:* dry transfer technique was used to place mechanically exfoliated $MoS_2$ mono-layers on pre-prepared hBN flakes. Electron-beam lithography (EBL) followed by reactive ion etching was used to pattern the $MoS_2$ channel. Electrodes were realized by EBL and e-beam evaporation and lift-off of Ti/Au (5nm/110nm) metals.

*SPEM characterization:* the X-ray beam energy used for these measurements was set to 701.3 eV. A mixture of Argon and $NO_x$ (1.7 ppm mol NO and 1.9 ppm mol $NO_2$) was introduced to the NAP-cell, at a pressure of $8.2 \times 10^{-4}$ mbar.

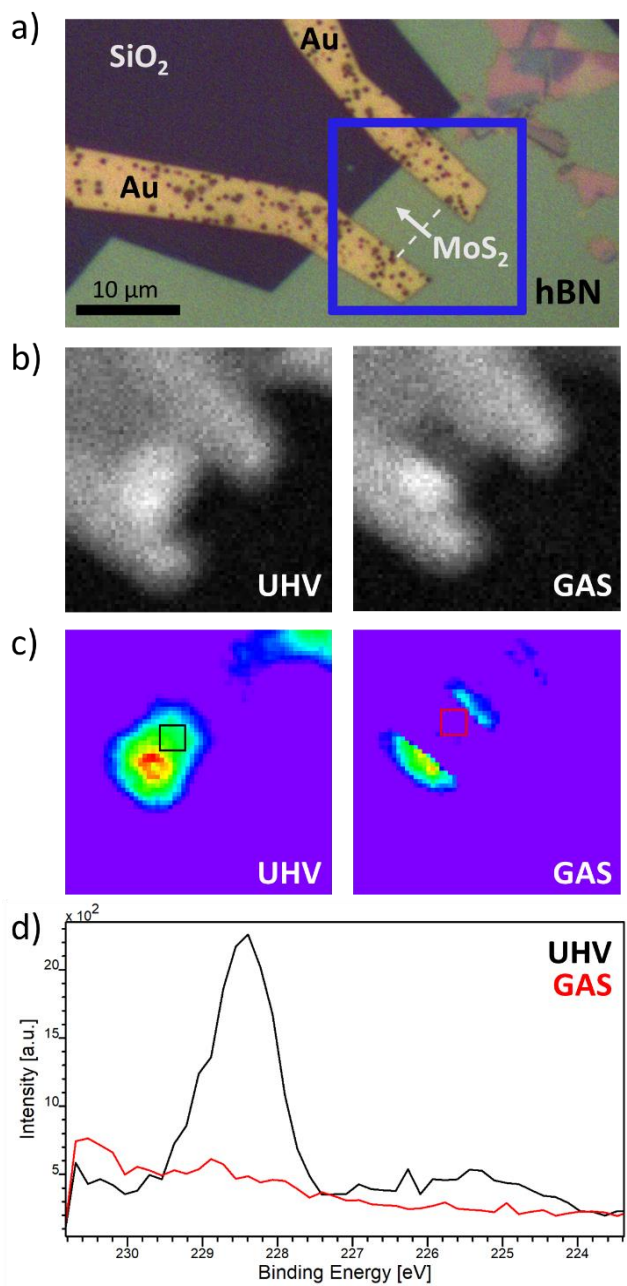

*Figure 12: (a) Optical image of the gas sensing device. (b) NAP-SPEM maps of the device in UHV (left) and NOx gas (right), corresponding to the area marked by the blue box in (a). (c) Map of the area of the Mo 3d$_{5/2}$ component obtained by fitting each spectrum contained in (b). (d) Spectra extracted from the boxes of corresponding colour displayed in (c).*

The maps were centred around KE 473 eV, corresponding to the Mo 3d$_{5/2}$ region. To extract local BE peak positions the data were first filtered by applying a 3x3 averaging filter to reduce the noise. Using an automated routine, peak deconvolution was performed at every pixel. New maps were created containing the total area of the deconvolution Mo 3d5/2 component for every pixel.

*Electrical characterization:* electrical transport properties were characterized using two-point configuration with a Keithley 2440 source meter. The gate voltage was applied using a Keithley 2450 instrument. All data were recorded by a purpose made LabVIEW program.

### 8.2 Results

The MoS$_2$-based FET was mounted into the NAP-cell with three independent electrical contacts for gate (V$_G$), source (V$_S$) and drain (V$_D$). Figure 12 (a) shows an optical image of the device, while Figure 12 (b) show SPEM maps centered at the Mo 3d$_{5/2}$ / S 2p signal, acquired before (left) and during exposure to NO$_x$ gas (right). In both maps, a bright spot can be seen close to the left side contact. This was caused by beam damage during initial focused beam measurements. Due to this beam damage, the current measured in UHV, prior to gas detection, was rather low. Nevertheless, the gas sensing device was still functioning, and a drop to zero conductivity was observed when NO$_x$ was introduced to the NAP-cell. Representative electrical transport properties of the device without local damage due to beam exposure can be found in Ref. [64].

The effect of the gas can be clearly seen as a decrease in intensity in the middle of the MoS$_2$ channel showed in Figure 12 (b) right. To investigate the effect of NO$_x$ gas on the MoS$_2$ channel more closely each spectrum contained in the SPEM maps was peak deconvoluted in an automated routine. Figure 12 (c) show the total area of the fitted Mo 3d$_{5/2}$ component, presented as maps. In UHV (left), the MoS$_2$ channel and the excess MoS$_2$ flake in the top right corner can be clearly distinguished from the parts of the device not containing MoS$_2$ (purple). By contrast, in NO$_x$ (right), the Mo 3d$_{5/2}$ component appears to have vanished from the middle of the MoS$_2$ channel and can only be found in the regions closest to the Au contacts. This is a result of the drop in conductivity induced by the gas, which cause the Mo 3d$_{5/2}$ component in the middle of the channel to be shifted out of the energy window of the SPEM map due to sample charging. Only the regions closest to the Au contacts are close enough to the contacts to avoid sample charging at such high gas pressures. Figure 12 (d) shows XPS spectra extracted from the same area of the MoS$_2$ channel in UHV and NO$_x$ gas, illustrated in Figure 12 (c) by black and red boxes, respectively. In UHV, the Mo 3d$_{5/2}$ component can be observed, while in NO$_x$ gas, it is no longer present and detailed spectra, not reported here, showed no indication of Mo oxide formation.

In summary, we have demonstrated how SPEM and NAP-CELL can be employed to investigate both *operando* chemical and electrical properties of 2D-materials based devices. The richness of the data offered by the chemical maps open the future possibility to investigate in detail the sensing mechanism of a single MoS$_2$ micrometric flake in a working gas sensor prototype.

## Conclusion

With the studies presented in this paper we demonstrate the wide possibility and flexibility offered by the combination of SPEM and NAP-CELL in a variety of different case studies and setups, from fundamental study, like metal oxidation dynamics, where reaction on specific sub-micrometric structure can be addressed, or following the intercalation of oxygen underneath a graphene coating, up to more applicative research where the behaviour of the active sensing material, in a working prototype of gas sensor, can be highlighted in realistic condition and at the relevant scale.

For research in surface science, microscopic and spectroscopic methods are a mainstay. The ability to have chemical identification at sub-micrometric scale, under *in-situ* and *operando* conditions, is very powerful in that it can yield more conclusive results when microscopy and spectroscopy are simultaneously combined in a single technique.

## Acknowledgements

The reported results present parts of different projects supported by Elettra-Sincrotrone Trieste SCpA, also the Calipso Programme is acknowledged for financial support under Grant Agreement 730872 from the EU Framework Programme for Research and Innovation HORIZON 2020